\documentstyle[aps]{revtex}
\begin{document}
\title{ Three-body Faddeev Calculation for $^{11}$Li with 
      Separable Potentials}
\author{ K. Ueta, H. Miyake}
\address{Instituto de F\'{\i}sica, Universidade de S\~ao Paulo,\\ 
CP 66318, \ 05315-970 S\~ao Paulo, S.P., Brasil}
\author{ G.W. Bund}
\address{Instituto de F\'{\i}sica Te\'{o}rica, Universidade Estadual 
Paulista,\\ Rua Pamplona 145, 01405-900 S\~{a}o Paulo, Brasil}
\date{\today}
\maketitle
\begin{abstract}
   The halo nucleus $^{11}$Li is treated as a three-body system 
consisting of an inert core of $^{9}$Li plus two valence neutrons.
The Faddeev equations are solved using separable potentials to
describe the two-body interactions, corresponding in the n-$^{9}$Li
subsystem to a p$_{1/2}$ resonance plus a virtual s-wave state. 
The experimental $^{11}$Li 
energy is taken as input and the $^{9}$Li transverse momentum 
distribution in $^{11}$Li is studied.
\end{abstract}
\bigskip
\draft
\pacs{PACS numbers: 21.45.+v, 21.60.-n, 11.80.Jy, 27.20.+n}
      Recent experiments with radioactive beams have permitted the study
of properties of nuclei close to the neutron drip line, that is, close to
the stability line for decay through neutron emission. These nuclei have 
been shown to possess neutron halos characterized by exceptionally
large radii and narrow  momentum distributions of the decay 
fragments in break-up experiments [1-2]. Among these nuclei
special attention has been paid to $^{11}$Li, a nucleus with a radius of 
about 3 fm and a two-neutron separation energy of only 0.3 MeV. In the 
case of $^{11}$Li grossly two lines of theoretical calculations have been 
followed; in one, the conventional shell model or Hartree-Fock approach is 
used[3], in the other a cluster model assuming a core of $^{9}$Li plus 
two neutrons is taken[4-8]. Because of the halo, shell model calculations 
require a very large single particle basis for the diagonalization of the 
hamiltonian. The cluster model seems particularly suited to the case of 
$^{11}$Li, considering the small 2n separation energy, the fact that 
$^{9}$Li is a normal nucleus with a neutron separation energy of 4 MeV 
and that $^{11}$Li is a Borromean nucleus, that is, no two-body subsystem 
of the three-body system does form a bound state. In the three-body model 
for $^{11}$Li, calculations are hampered by a lack of information with 
regard to the n-$^{9}$Li interaction. Several calculations[4-7,24] were 
performed using local potentials for this interaction and there is also 
one calculation using separable potentials[8]. The parameters of the 
potentials were adjusted to produce a n-$^{9}$Li resonance,
which has most frequently been assumed in the p$_{1/2}$ channel, 
although the experimental data are not conclusive. According to Wilcox 
et al.[9] a resonance occurs  at   $\,0.80\,\pm\,  0.25$~MeV. Following 
more recent work[10] there is a 2$^{+}$ resonance situated at 0.42~MeV and 
a 1$^+$ \ state at 0.80~MeV. In addition one has now strong evidence for 
an enhancement of the production of $^{10}$Li near threshold in reactions 
involving $^{11}$Be, $^{11}$Li and $^{11}$B[11,12]. This is interpreted 
as due to an intruder virtual s-wave state of the n-$^9$Li system near 
zero energy corresponding to a scattering length of -20fm or less [12]. 
A three-body study [13] indicated that this barely unbound state in 
$^{10}$Li is able to explain the extra narrowness of the momentum 
distribution of $^9$Li in the fragmentation of $^{11}$Li. 
   With a s-wave scattering length of -44 fm and a p$_{1/2}$ resonance
energy  of 0.35 MeV, the  2n separation energy in $^{11}$Li and the
momentum distribution of the $^9$Li fragment are fitted in Ref.[13] 
assuming the sudden approximation[14] and neglecting the final state 
interaction in the break-up reaction.
   It is our purpose here, to make a similar fit applying the 
three-body model with separable potentials developed earlier[15]
 and which proved to be effective in the description of the structure
of $^{18}$O and $^{18}$F as well as $^{16}$O(d,p) stripping.\\
\indent 
    We consider the halo nucleus $^{11}$Li as a three-body system 
consisting of a $^{9}$Li core (particle 3), which stays inert, plus 
two valence neutrons (particles 1 and 2). The valence neutrons couple
with the orbital angular momentum of the core to total angular
momentum and parity J$^{\pi}$=0$^{+}$, the spin and parity of
$^{11}$Li being then due to the value $\frac{3}{2}^{-}$ of the $^9$Li
core. The neutron-$^9$Li system is assumed to have a p$_{1/2}$ 
resonance of width $\Gamma$ = 0.15$\pm$0.07 MeV at an energy 
E$_r$=0.42$\pm$0.05 MeV [10] and also a s$_{1/2}$ virtual state close 
to zero energy [12,16]. We must also take into account
the Pauli principle which does not allow a valence neutron to occupy
the 1s$_{1/2}$ and the 1p$_{3/2}$ single particle states which are 
already filled in the core. To fix the energies 
$\epsilon_{1s_{1/2}}$ and $\epsilon_{1p_{3/2}}$ of these states, we
proceed as follows. First by doing an interpolation between the 
experimental value -4.053 MeV of $\epsilon_{1p_{3/2}}$ for A=9 [17]
and the values in Fig.2-30 of Bohr-Mottelson's book [18] we obtain
$\epsilon_{1p_{3/2}} \simeq -7$ MeV for the A=10 system. To obtain
$\epsilon_{1s_{1/2}}$ we suppose that the separation between the 
1s$_{1/2}$ level and the centroid of the levels 1p$_{3/2}$ and
1p$_{1/2}$ is $\hbar\,\omega$ and use the prescription 
$\hbar\,\omega \,=\, 45 \; A^{-1/3} - 25 \; A^{-2/3} \,{\rm MeV}$,
which is appropriate for light nuclei [19]. For A=10, one has 
$\hbar\,\omega$=15.501 MeV and $\epsilon_{1s_{1/2}}$ results equal to
-20.028 MeV. To account for the Pauli blocking of the states 
1s$_{1/2}$ and 1p$_{3/2}$, we use the projection method of Kukulin
[20]. From now on, we consider $\hbar$=1.\\
\indent
   To describe the neutron-$^9$Li interaction we use a separable 
potential which acts on the s$_{1/2}$, p$_{3/2}$ and p$_{1/2}$
waves:
\begin{eqnarray}
\langle{\bf P}_i|V_{i}|{\bf P}_i' \rangle\,& = &\,\sum_{l\,j\,\alpha}
 -\frac{\Lambda_{lj}^{\left(\alpha\right)}}{2m}
   \;v_{lj}^{\left(\alpha\right)}(P_i)\;
   v_{lj}^{\left(\alpha\right)}(P_i')\;\nonumber\\
&   &\times\sum_\mu \;\langle\hat{\bf P}_i | y_{l\,j\,\mu}\rangle\;\langle
         y_{l\,j\,\mu}\, |\hat{\bf P}_i'\rangle \quad \, (i=1,2) 
       \quad , 
\end{eqnarray}
\noindent 
where l\,j = 0\,$\frac{1}{2}$, 1\,$\frac{3}{2}$ and
1\,$\frac{1}{2}$ , $\,m\,$ is the reduced mass of the n-$^9$Li
system  ($\,m\,$=$\frac{9}{10}$M, M being the nucleon  mass) and
${\bf P}_i$ is the momentum of neutron i with respect to $^9$Li.
The s$_{1/2}$ potential is a three-term ($\alpha$=1,2,3) potential, 
with the form factors chosen as
\begin{equation}
v_{0\frac{1}{2}}^{\left(1\right)}(q)\,=\,
     (q^2+\alpha^2_{0\frac{1}{2}})\,
      \exp\left(-\beta_0^2 q^2/2\right)\quad ,
\end{equation}
\begin{equation}
v_{0\frac{1}{2}}^{\left(2\right)}(q)\,=\,
     (\frac{3}{2}-\beta_0^2 q^2)\,
      \exp\left(-\beta_0^2 q^2/2\right)\quad ,
\end{equation}
\begin{equation}
v_{0\frac{1}{2}}^{\left(3\right)}(q)\,=\,
     \exp\left(-\beta_0^2 q^2/2\right)\quad .
\end{equation}
\noindent
The first term is chosen in such a way that, alone, it reproduces the
 1s$_{1/2}$ state. Taking for 
$\alpha_{0\frac{1}{2}}$ and
$\Lambda_{0\frac{1}{2}}^{\left(1\right)}$ the special values
\begin{equation} 
\alpha_{0\frac{1}{2}} = \sqrt{\,2m|\epsilon_{1s_{1/2}}|\,} \,
\end{equation}
 and 
\begin{equation}
\Lambda_{0\frac{1}{2}}^{\left(1\right)} \,=\,\left[
\int^\infty_0 dq\, q^2\,\frac{[v_{0\frac{1}{2}}^{\left(1\right)}(q)]^2}{q^2
   - 2m\epsilon_{1s_{1/2}}}\right]^{-1} \quad,
\end{equation}
\noindent
we get a bound state of energy $\epsilon_{1s_{1/2}}$ and wave 
function
\begin{equation}
  \Phi_{0\frac{1}{2}\mu}({\bf q}) =
     N_{0\frac{1}{2}} \,\exp\left(-\beta_0^2 q^2/2\right) 
     y_{0\frac{1}{2}\mu}(\hat{\bf q})\quad,
\end{equation}
\noindent
which is precisely the 1s$_{1/2}$ oscillator function in momentum 
space if we set $\,\beta_0 \,= \,1/\sqrt{m\omega}\,$.
\noindent From the values chosen for $\epsilon_{1s_{1/2}}$ and 
$\omega$, we obtain $\alpha_{0\frac{1}{2}}$=0.932 fm$^{-1}$,
$\Lambda_{0\frac{1}{2}}^{\left(1\right)}$=8.417 fm$^5$ and
$\beta_0$=1.724 fm.\\
\indent
   The addition of the second term to the s$_{1/2}$ potential
does not affect the bound state generated by the first term. This
is a consequence of the orthogonality relation
\begin{equation}
\int^\infty_0 dq\, q^2\,[\exp\left(-\beta_0^2 q^2/2\right)]
\,[(\frac{3}{2}-\beta_0^2 q^2)\, \exp\left(-\beta_0^2 q^2/2\right)]
=0\quad .
\end{equation}
\noindent However, the scattering states 
are affected and the two terms together can give rise to a virtual 
s$_{1/2}$ state. 
For $\Lambda_{0\frac{1}{2}}^{\left(2\right)}$=2.696 fm, we obtain a 
virtual state placed at an energy $\epsilon_v$=-40 keV on the second 
Riemann sheet. The corresponding scattering length is 
a$_{s_{1/2}}$=-20 fm.  We remark here that the virtual state has a
2s$_{1/2}$ character, since by increasing slightly the strength of
the second term ($\alpha$=2) in the s$_{1/2}$ wave potential, it 
becomes a bound state with two nodes.\\
\indent
    The third term of the s$_{1/2}$ potential is the projection operator
for the forbidden 1s$_{1/2}$ state, constructed according to the 
prescription of Kukulin[20]. The corresponding form factor (Eq.(4)), 
being proportional to the wave function of the bound state
produced by the first two terms, is orthogonal to the scattering 
states generated by these terms.  Therefore the third term does not 
affect the scattering and the virtual state remains unchanged. It is 
not so for the 1s$_{1/2}$ bound state. Although the wave function 
given by expression (7) remains an eigenfunction for the three-term 
potential, the corresponding energy is affected. By considering the 
third term repulsive 
($\Lambda_{0\frac{1}{2}}^{\left(3\right)}\,<\,0$ ), we remove the bound 
state to the continuum part of the spectrum (therefore, it becomes 
a continuum bound state) and, by making
$\Lambda_{0\frac{1}{2}}^{\left(3\right)} \rightarrow -\infty$, the
forbidden 1s$_{1/2}$ state is projected out.\\ 
\indent
    For the p$_{3/2}$ potential, we consider a two-term potential
with form factors
\begin{equation}
v_{1\frac{3}{2}}^{\left(1\right)}(q)\,=\,
      q\,(q^2+\alpha^2_{1\frac{3}{2}})\,
      \exp\left(-\beta_1^2 q^2/2\right)\quad ,
\end{equation}
\begin{equation}
v_{1\frac{3}{2}}^{\left(2\right)}(q)\,=\,
     q\,\exp\left(-\beta_1^2 q^2/2\right)\quad .
\end{equation}
\noindent
The first term of the potential is chosen so as to reproduce the 
1p$_{3/2}$ bound state. With the choice
$\alpha_{1\frac{3}{2}}$ = $\sqrt{\,\,2m\,|\epsilon_{1p_{3/2}}|\,}$ 
and 
\begin{equation}
\Lambda_{1\frac{3}{2}}^{\left(1\right)} \,=\,\left[
\int^\infty_0 dq\, q^2\,\frac{[v_{1\frac{3}{2}}^{\left(1\right)}(q)]^2}
{q^2- 2m\epsilon_{1p_{3/2}}} \right]^{-1} \quad ,
\end{equation}
\noindent
the potential produces a bound state of energy $\epsilon_{1p_{3/2}}$
 and wave function identical to the 1p$_{3/2}$ harmonic oscillator 
wave function
\begin{equation}
  \Phi_{1\frac{3}{2}\mu}({\bf q}) =
     N_{1\frac{3}{2}} \,q\,\exp\left(-\beta_1^2q^2/2\right) 
     y_{1\frac{3}{2}\mu}(\hat{\bf q})\quad .
\end{equation}
For the choice $\beta_1$=$\beta_0$=1.724 fm and
$\epsilon_{1p_{3/2}}$=-7 MeV, we obtain 
$\alpha_{1\frac{3}{2}}$=0.551 fm$^{-1}$ and
 $\Lambda_{1\frac{3}{2}}^{\left(1\right)}$=20.018 fm$^7$.\\
\indent
    The second term in the p$_{3/2}$ potential, with 
 $\Lambda_{1\frac{3}{2}}^{\left(2\right)} \rightarrow -\infty$,
is nothing but the projection operator which projects out the 
forbidden 1p$_{3/2}$ state, the scattering states remaining the 
same as those produced by the first term alone. Thus, the p$_{3/2}$
phase shift is dominated by the occupied 1p$_{3/2}$ bound state,
there being no resonances as it might occur in the case of a local
potential.\\
\indent
    The p$_{1/2}$ potential is taken as a one term potential
with the form factor given by
\begin{equation}
v_{1\frac{1}{2}}^{\left(1\right)}(q)\,=\,
      q\,(q^2+\alpha^2_{1\frac{1}{2}})\,
      \exp\left(-\beta_1^2 q^2/2\right)\quad .
\end{equation}
Thus, $v_{1\frac{1}{2}}^{\left(1\right)}$ is of the same form as
$v_{1\frac{3}{2}}^{\left(1\right)}$. For simplicity we have also assumed 
the same parameter $\beta_1$ in the exponential part of the p$_{1/2}$
 and p$_{3/2}$ form factors.
We initially make the choice
$\alpha_{1\frac{1}{2}}= \alpha_{1\frac{3}{2}}$, thus getting
$\alpha_{1\frac{1}{2}}$=0.551 fm$^{-1}$. Using this value and the 
condition that the p$_{1/2}$ resonance occurs at 0.42 MeV, we obtain
$\Lambda_{1\frac{1}{2}}^{\left(1\right)}$=13.535 fm$^7$. The width
of the resonance turns out to be 0.12 MeV and compares with the
experimental value mentioned before.\\
\indent
    Regarding the interaction between the valence neutrons, we 
assume a free  nucleon-nucleon interaction considering the low nucleon 
density in the region of the halo. For the 0$^{+}$ state which we are 
considering, if one assumes pure single particle harmonic oscillator
states, the (p$_{1/2}$)$^2$ configuration is a superposition of
both the $^1$S$_0$ and $^3$P$_1$ states (33 and 67 percent 
respectively) of the n-n subsystem, while only $^1$S$_0$ appears in 
the (s$_{1/2}$)$^2$ configuration. Calculations by other authors 
(for instance, Ref.[13]) have shown that the $^3$P$_1$ potential, 
which is repulsive, changes the 2n separation energy by at least 
50$\%$. We therefore use a potential which acts in both $^{1}$S$_0$ 
and $^3$P$_{1}$ channels, and take it separable,
\begin{eqnarray}
\langle{\bf p}|V_{12}|{\bf p}' \rangle\,& = &\,\sum_{\lambda\,S\,I} 
-\frac{\Lambda_{\lambda\,S\,I}}{\rm M}
\;v_{\lambda\,S\, I}\left(p\right)\;
  v_{\lambda\,S\, I}\left(p'\right)\;\nonumber\\
&    & \times\sum_{M_{I}} \;\langle\hat{\bf p} |
 y_{\lambda\,S\,}^{I\,M_{I}}\rangle\;\langle
 y_{\lambda\,S\,}^{I\,M_{I}}\, |\hat{\bf p}'\rangle \quad ,
\end{eqnarray}
where {\bf p} is the relative momentum between the two neutrons, 
($\lambda\,S\,I$)=(000) and (111) and the form factors 
are of the Yamaguchi type:
\begin{equation}
 \,v_{000}(p)\,=\,\left[ p^2 + \alpha^2_{000}\right] ^{-1} ,
\end{equation}
\begin{equation}
 \,v_{111}(p)\,=\,p\left[ p^2 + \alpha^2_{111}\right] ^{-2}.
\end{equation}
  Using the values a$_{{^1}{\rm S}_0}$=-17 fm and  
r$_{{^1}{\rm S}_0}$=2.84 fm for the neutron-neutron scattering length
 and effective range in 
the $^1$S$_0$ channel [21] and a$_{{^3}{\rm P}_1}$=2.2 fm$^{3}$ and
 r$_{{^3}{\rm P}_1}$=-8.0 fm$^{-1}$ for the $^3$P$_1$ channel [22], we
 fix the parameters of the V$_{12}$ potential as 
 $\,\Lambda_{000}\,=\, 1.662$ fm$^{-3}\,$ ,
$\,\alpha_{000}  =1.130$ fm$^{-1}\,$,
$\,\Lambda_{111} =-0.078 $ fm$^{-5}\,$ and
$\,\alpha_{111}\,=\,0.693$ fm$^{-1}\,$. The negative value of
$\,\Lambda_{111}$ means that the $^3$P$_1$ term is repulsive.\\
\indent
   The proposed interactions are used as input 
in the homogeneous Faddeev equations 
\begin{eqnarray}
\Psi^{(1)} &=& G_0\; T_1 \,\left(\Psi^{(2)} + 
       \Psi^{(3)}\right)\quad ,\nonumber \\ [0.3cm]
\Psi^{(2)} &=& G_0\; T_2 \,\left(\Psi^{(3)} + 
       \Psi^{(1)}\right)\quad , \\ [0.3cm]
\Psi^{(3)} &=& G_0\; T_{12} \,\left(\Psi^{(1)} + 
       \Psi^{(2)}\right)\quad ,\nonumber 
\end{eqnarray}
where $\,\Psi^{(1)}\,$, $\,\Psi^{(2)}\,$ and $\,\Psi^{(3)}\,$ 
are the Faddeev components of the total wave function,
\begin{equation}
\,\Psi = \Psi^{(1)} + \Psi^{(2)} + \Psi^{(3)}\;\quad , 
\end{equation}
\noindent 
$\,G_0 = \left(E - H_0\right)^{-1}\,$ is the Green's function
and $\,T_1, T_2\,$ and $\,T_{12}\,$ are the $T$-matrices
corresponding to the potentials $\,V_1, V_2\,$ and $\,V_{12}\,$
respectively.  After performing the angular momentum 
decomposition, we end up with a homogeneous system of coupled 
integral equations in one variable. The equations are then transformed
into a set of homogeneous algebraic equations using the Gauss quadrature 
method to approximate each integral by a finite sum. The zero of the 
determinant of this system of equations gives the 
separation energy, S$_{2n}$, of the two valence neutrons. 
From the corresponding three-body wave function, we calculate the
momentum distributions. We restrict ourselves to the calculation
of the transverse momentum distribution of the $^9$Li core in $^{11}$Li.
We should mention that, in our model, the transverse and the parallel
momentum distributions turn out to be identical. An approximate equality 
has indeed been verified experimentally for the $^9$Li fragment in the 
break-up of $^{11}$Li [23]. Corrections due to final state interactions 
in the case of the momentum distribution of the core fragment  should be 
small according to reference [24].\\
\indent 
    For the chosen values of the parameters, we obtain the value 
S$_{2n}$=0.293 MeV which agrees with the (average) experimental value 
0.294$\pm$0.030 MeV, reported in reference [25]. However, the 
width of the transverse momentum distribution of $^9$Li turns out
 to be too large, as shown in the figure (dotted line). The 
experimental data are from Kobayashi et al. [26] (actually, the 
experimental points shown in the figure are taken from Fig. 2 of 
reference [6]).  We here mention that, if the $^3$P$_1$ component of 
the n-n interaction is suppressed the value of S$_{2n}$ duplicates,
becoming equal to 0.597 MeV.\\
\indent
   The result may be improved by taking for $\beta_0$ a value smaller
than the one given by the prescription
$\beta_0 = 1/\sqrt{m\omega}$ = 1.724fm.
We made a new calculation in which the s$_{1/2}$ and p$_{1/2}$ 
potentials are modified as follows.
Assuming $\beta_0$=1.5fm and maintaining the previous values 
$\epsilon_{1s_{1/2}}$ = -20.028MeV and 
$\alpha_{0\frac{1}{2}}$=0.932 fm$^{-1}$, we determine
$\Lambda^{(1)}_{0,\frac{1}{2}}$ to be 4.959 fm$^5$. By
setting $\Lambda^{(2)}_{0,\frac{1}{2}}$ = 2.219 fm the virtual state
is positioned at $\epsilon_{v}$=-40keV, the corresponding 
scattering length being a$_{s_{1/2}}$=-20fm.
 To determine the p$_{1/2}$ potential, we keep $\beta_1$ unchanged
and adjust $\Lambda^{(1)}_{1,\frac{1}{2}}$ and 
$\alpha_{1,\frac{1}{2}}$ with the condition that the calculated p$_{1/2}$ 
resonance energy occurs at 0.42 MeV, the resonance width $\Gamma$ 
remains inside the range 0.15$\pm$0.07 MeV and, in addition, that 
the calculated 2n separation energy agrees with the experimental value  
0.294MeV. The parameters of the p$_{3/2}$ interaction are the same
as in the previous calculation. 
 The corresponding transverse momentum distribution is given by the 
dashed line of the figure. The half width at half maximum(HWHM) 
is 40 MeV/c and the three-body state is a superposition mainly of  
$^1$S$_0$ (47$\%$) and $^3$P$_1$ (48$\%$) configurations of the n-n
subsystem.\\
\indent
Repeating the previous calculation, with a still smaller value,  
$\beta_0$=1.4 fm, we get 35 MeV/c for the HWHM. The agreement with 
experiment is very good (solid line of the figure).
A similar result for the transverse momentum distribution is also
obtained taking $\beta_0$=1.5 fm, and changing the scattering length
to a$_{s_{1/2}}$=-40 fm (dot-dashed line). In Table 1 we summarize the 
parameters of the two-body s$_{1/2}$ and p$_{1/2}$ potentials used 
in the curves I-IV and in Table 2 the corresponding three-body results. 
The increase of the n-n $^1$S$_0$ contribution in the cases II-IV, 
points to an increase of the contribution of the (s$_{1/2}$)$^2$ 
configuration to the three-body wave function. \\
\indent
    Our calculation shows that relevant $^{11}$Li data may be fitted by a 
three-body model using simple separable potentials with parameters
adjusted to two body data. The fact that one needs for the parameter 
$\beta_0$ a value not in accordance with the prescription
$\,\beta_0 \,= \,1/\sqrt{m\omega}\,$ indicates that the effective
neutron-$^9$Li potential may differ appreciably from the usual 
single-particle potential. The use of a local neutron-$^9$Li potential
should corroborate this conclusion. In fact, one can verify, by using for 
instance a simple square-well potential, that in order to obtain an 
intruder 2s$_{1/2}$ state near zero energy a much deeper potential than 
usually is needed. Such odd potentials may indicate that a more 
detailed description, which does not consider the core as a 
structureless object, is required.\\
The numerical calculations were performed at LCCA-USP.
\bigskip
\newpage
\begin{table}
\caption {Parameters of the two-body s$_{1/2}$ and p$_{1/2}$ 
potential corresponding to the curves I-IV presented in Fig. 1.
These parameters correspond to $\epsilon_{1s_{1/2}}$ = -20.028MeV
and E$_r$=0.42MeV. For the p$_{3/2}$ potential we use 
$\beta_1$=1.724fm, $\Lambda^{(1)}_{1,\frac{3}{2}}$=20.018fm$^{-1}$, 
$\alpha_{1,\frac{3}{2}}$=0.551fm$^{-1}$, corresponding to 
$\epsilon_{1p_{3/2}}$=-7MeV. The width $\Gamma$ of the p$_{1/2}$ 
resonance and the scattering length a$_{s_{1/2}}$ are also given.}
\begin{tabular}{llllllllll}
 &$\beta_0$&$\alpha_{0,\frac{1}{2}}$       &
           $\Lambda^{(1)}_{0,\frac{1}{2}}$&
           $\Lambda^{(2)}_{0,\frac{1}{2}}$&
 $\beta_1$&$\alpha_{1,\frac{1}{2}}$       &
           $\Lambda^{(1)}_{1,\frac{1}{2}}$& 
           $\Gamma$                       &
           a$_{s_{1/2}}$\\ 
 &(fm)&(fm$^{-1}$)&(fm$^5$)&(fm)&(fm)&(fm$^{-1}$)&(fm$^7$)&(MeV)&(fm)\\
\tableline
I  &1.724&0.932&8.417&2.696&1.724&0.551&13.535& 0.12 & -20 \\
II &1.5  &0.932&4.959&2.219&1.724&0.656&10.628& 0.16 & -20 \\
III&1.4  &0.932&3.788&2.008&1.724&0.746& 8.502& 0.20 & -20 \\
IV &1.5  &0.932&4.959&2.304&1.724&0.709& 9.328& 0.18 & -40 \\
\end{tabular}
\label{Table 1}
\end{table}
\bigskip
\begin{table}
\caption {Three-body results corresponding to the parameters 
listed in Table 1. S$_{2n}$ is the 2n separation energy from $^{11}$Li
and $\gamma$ is the HWHM of the calculated momentum distribution
of the $^9$Li core in $^{11}$Li. The last four columns give the fractional 
admixture of n-n states in the $^{11}$Li wave function.}
\begin{tabular}{lllllll}
 &S$_{2n}$ &$\gamma$&$^1$S$_0$&$^3$P$_1$&$^1$D$_2$&$^3$F$_3$\\
 &(MeV)    &(MeV/c) &         &         &         &         \\
\tableline
I  &0.293& 82 & 0.38 & 0.59 & 0.01 & 0.02\\
II &0.294& 40 & 0.47 & 0.48 & 0.02 & 0.02\\
III&0.294& 35 & 0.54 & 0.41 & 0.03 & 0.02\\
IV &0.294& 35 & 0.55 & 0.40 & 0.04 & 0.02\\
\end{tabular}
\label{Table 2}
\end{table}
\bigskip
\begin{figure}
\caption{Transverse momentum distribution of $^9$Li in the 
$^{11}$Li. The squares and circles are experimental data[26] corresponding 
to P$_{\perp}$$<$0 and P$_{\perp}$$>$0 respectively. Curve (I) corresponds
 to $\beta_0$=$\beta_1$=1.724fm and scattering length 
a$_{s_{1/2}}$=-20fm, (II) to $\beta_0$=1.5fm, $\beta_1$=1.724fm,  
a$_{s_{1/2}}$=-20fm, (III) to $\beta_0$=1.4fm, $\beta_1$=1.724fm, 
a$_{s_{1/2}}$=-20fm and (IV) to $\beta_0$=1.5fm, $\beta_1$=1.724fm, 
a$_{s_{1/2}}$=-40fm.}
\label{Fig.1}
\end{figure}
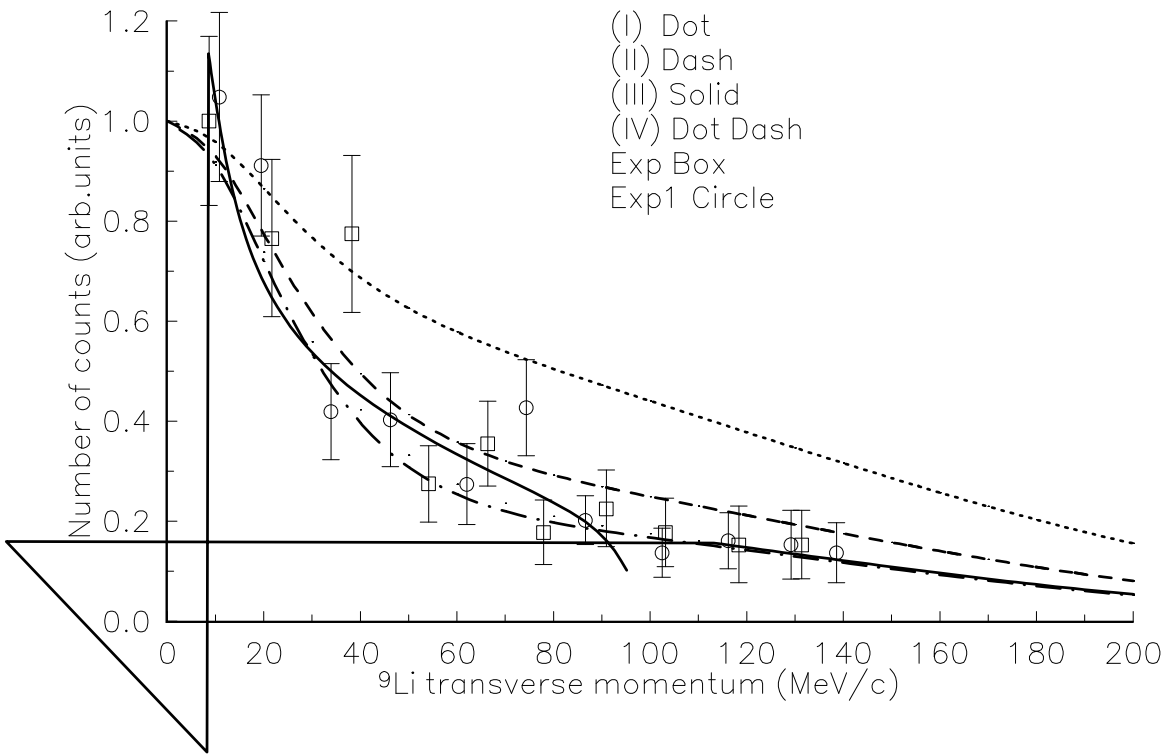
\end{document}